\documentclass[revtex4,numberedappendix,twocolappendix]{emulateapj} 
\slugcomment{{\sc Accepted to ApJ:} November 15, 2017} 
\usepackage{natbib,amsmath,booktabs,apjfonts,threeparttable,multirow}
\usepackage{natbib,ifsym}
\usepackage{subfigure}
\usepackage{graphicx}
\usepackage{tabularx}
\usepackage[colorlinks=true,linkcolor=blue,citecolor=blue,urlcolor=blue]{hyperref}
\usepackage{epstopdf}
\usepackage{enumitem}
\usepackage{color}
\usepackage{pdfpages}

\begin{document}
	
\title{Comparison of the X-ray emission from Tidal Disruption Events with those of Active Galactic Nuclei}

\author{Katie Auchettl\altaffilmark{1,2},  Enrico Ramirez-Ruiz\altaffilmark{3} and James Guillochon\altaffilmark{4}}

\altaffiltext{1}{Center for Cosmology and Astro-Particle Physics, The Ohio State University, 191 West Woodruff Avenue, Columbus, OH 43210, USA}
\altaffiltext{2}{Department of Physics, The Ohio State University, 191 W. Woodruff Avenue, Columbus, OH 43210, USA}
\altaffiltext{3}{Department of Astronomy and Astrophysics, University of California, Santa Cruz, CA 95064, USA}
\altaffiltext{4}{Harvard-Smithsonian Center for Astrophysics, 60 Garden St., Cambridge, MA 02138 USA}

\begin{abstract}
	One of the main challenges of current tidal disruption events (TDEs) studies is that emission arising from AGN activity may potentially mimic the expected X-ray emission of a TDE. Here we compare the X-ray properties of TDEs and AGN to determine a set of characteristics which would allow us to discriminate between flares arising from these two objects. We find that at peak, TDEs are brighter than AGN found at similar redshifts. However, compared to preflare upperlimits, highly variable AGN can produce flares of a similar order of magnitude as those seen from X-ray TDEs. Nevertheless, TDEs decay significantly more monotonically, and their emission exhibits little variation in spectral hardness as a function of time. We also find that X-ray TDEs are less absorbed, and their emission is much softer than the emission detected from AGN found at similar redshifts. We derive the X-ray luminosity function (LF) for X-ray TDEs using the events from \citet{2016arXiv161102291A}. Interestingly, our X-ray LF matches closely the theoretically derived LF by \citet{2006ApJ...652..120M} which assumes a higher TDE rate currently estimated from observations. Using our results and the results of \citet{2016MNRAS.455..859S}, we estimate a TDE rate of $(0.7-4.7)\times10^{-4}$ yr$^{-1}$ per galaxy, higher than current observational estimates. We find that TDEs can contribute significantly to the LF of AGN for $z\lesssim0.4$, while there is no evidence that TDEs influence the growth of $10^{6-7}M_{\odot}$ BHs. However, BHs $<10^{6}M_{\odot}$ can grow from TDEs arising from super-Eddington accretion without contributing significantly to the observed AGN LF at $z=0$.

\end{abstract}

\section{Introduction}
When a star passes within the tidal radius of a supermassive black hole (SMBH), it will be partially \citep[e.g.,][]{2013ApJ...767...25G, 2014ApJ...783...23G} or completely \citep{1975Natur.254..295H, 1982ApJ...262..120L, 1988Natur.333..523R} disrupted by the tidal forces of the BH. Approximately half of the debris from this tidal disruption event (TDE) will fall back onto the BH and form a viscous accretion disc. During a full disruption of a main-sequence \citep{2013ApJ...767...25G} or giant branch star \citep{2012ApJ...757..134M}; or during the partial disruption of a main-sequence star, the eventual accretion of this material will produce a luminous flare \citep{1982ApJ...262..120L, 1988Natur.333..523R, 1989ApJ...346L..13E, 1989Natur.340..595P} that is short lived relative to the expected rate of occurrence\footnote{For a  $M_{BH}\lesssim10^{7.5}M_{\odot}$, one expects $\sim10^{-4}$ TDEs per year.} \citep{1999MNRAS.309..447M, 2004ApJ...600..149W, 2016MNRAS.455..859S}.

A large fraction of the luminosity produced during a TDE will fall within the soft X-ray band \citep{1999ApJ...514..180U}, dominating the fainter, more permanent X-ray emission of its host. As a consequence, searching for TDEs in X-ray wavelengths has lead to the identification of a number of candidate events \citep[see e.g.,][and references therewithin]{2016arXiv161102291A}. 
However, establishing that a transient X-ray flare arises from a TDE is complicated by the fact that Active Galactic Nuclei (AGN), which are usually found in a low-luminosity, quiescent state \citep{2008ARA&A..46..475H}, can also produce highly luminous flares over short duty cycles from slim disk accretion instabilities arising from their long lived accretion disks \citep[e.g., ][]{1991PASJ...43..147H}. 

In fact, this ambiguity between TDEs and AGN is not just limited to their observed X-ray light-curves. Recently, \citet{2015MNRAS.452...69M} suggested that TDEs may account for 1-10\% of all detected AGN in wide-field, multi-wavelength snapshot surveys \citep[see review by][and references therewithin]{2015A&ARv..23....1B}, while e.g., \citet{2006ApJ...652..120M} and \citet{2013ApJ...777..133M} have showed theoretically that TDEs can contribute significantly to the quiescent luminosity of AGN. In addition, the perturbing presence of a long-lived accretion disc like those seen in AGN can actually enhance the rate of TDEs \citep{2007A&A...470...11K}.

Due to the relatively young nature of the field associated with observationally characterising TDEs in the X-rays, the current ambiguity associated with classifying an X-ray transient coincident with the centre of an inactive galaxy as an AGN or a TDE might not be so surprising. Currently, the rate at which we observe these events, and the expected theoretical rate of TDEs in inactive galaxies differ by nearly an order of magnitude \citep[see][ and references therewithin]{2016MNRAS.461..371K}, while recently, \citet{2016arXiv161102291A} showed that out of nearly 70 potential TDE candidates suggested in the literature, only a handful of well-characterized events exhibited properties expected of an X-ray TDE. 

Of the events that were re-classified by \citet{2016arXiv161102291A} to be ``not a TDE'', a number were found to have properties more consistent with being AGN\footnote{Nearly $20$\% of the TDE candidates that were analysed in \citet{2016arXiv161102291A} turned out to be more consistent with emission arising from an AGN rather than a TDE.}, and the true nature of these events only became clear after significant multi-wavelength follow-up and/or obtaining long-term X-ray light curves that span over multiple decades. As such, a major challenge of current X-ray TDE studies is to efficiently and cleanly select candidates without requiring significant resources and observational time to follow-up each potential candidate. This highlights that our understanding of the differences between the X-ray properties of TDEs compared to that of AGN is currently unsatisfactory, and thus dramatically limits our abilities to correctly classify an X-ray transient event a TDE.

As such, we attempt to quantify the observed differences in the X-ray flare emission arising from a TDE or AGN, with the goal of providing a set of characteristics which can help improve the credibility of a TDE classification for an unknown flare based solely on its X-ray emission. Using the TDE candidates classified as an ``X-ray TDE'' or a ``likely X-ray TDE'' in \citet{2016arXiv161102291A} (see Table 2 of this paper), we compare and contrast the X-ray properties of these events with the X-ray properties of AGN detected in both extragalactic  X-ray surveys and detailed follow-up observations of individual sources. In Sections 2--5 we compare the various properties our TDE sample to those of AGN and discuss the implications of each of studies as we go, while in Section 4 we summarise the our main findings.

\section{Brightness as a function of redshift}

\begin{figure}[t]
	\begin{center}
		\includegraphics[width=1.05\columnwidth]{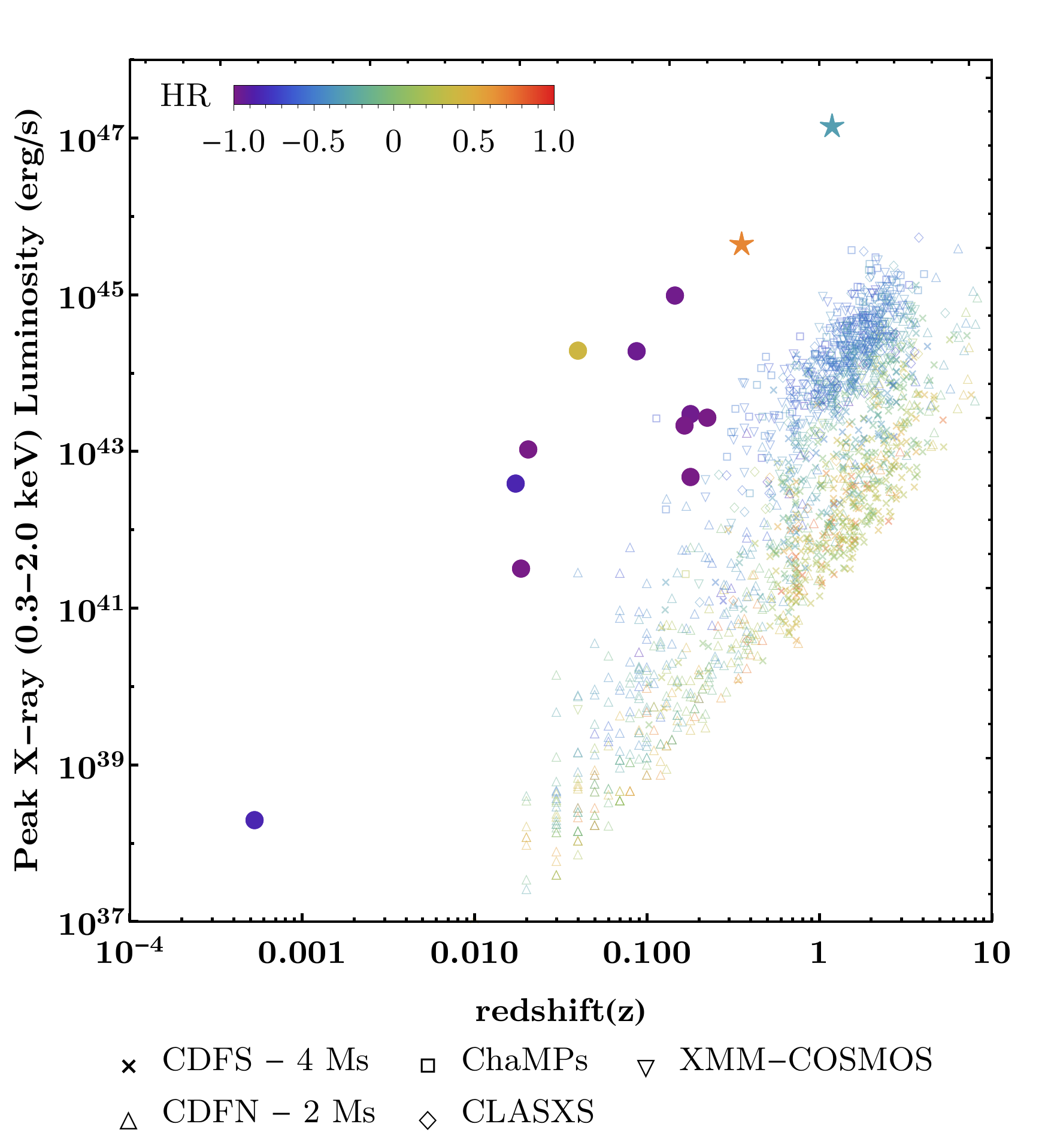}
		\caption{Soft (0.3-2.0 keV) X-ray luminosity plotted as a function redshift. Here, we have plotted the peak X-ray luminosities of the X-ray TDE and likely X-ray TDE candidates classified by \citet{2016arXiv161102291A}, and the soft X-ray luminosities of AGNs detect using the 4Ms Chandra Deep Field South ($\times$), 2Ms Chandra Deep Field North ($\triangle$), Chandra Multiwavelength Project (ChaMPs) ($\square$), Chandra Large Area Synoptic X-ray Survey (CLASXS) ($\diamond$) and the XMM-COSMOS survey ($\triangledown$). See text for references to these surveys. The colours of each symbol represents how hard or soft (i.e., the hardness ratio, HR) the detected X-ray emission is arising from these objects. A source is considered soft if it has a $HR=-1$, while it is considered hard if it has a $HR=-1$. Listed in decreasing redshift we have plotted jetted TDEs \textit{Swift J2058+05} and \textit{Swift J1644+57} as a star ($\bigstar$), while those plotted with a filled circle ($\bullet$) are either classified as non-jetted or have no classification.\label{lumredvshr}}
	\end{center}
\end{figure}

Deep extragalactic surveys allow us to probe various astrophysical populations at high redshift. In particular, the study of AGN at various cosmological distances and how they evolve with time provides key information for us to better understand the accretion history of SMBH. It has been shown that AGN are most active between a redshift ($z$) of $1-3$ \citep[see the X-ray luminosity functions derived by e.g.,][]{2015MNRAS.451.1892A}, however it is well known that there is a significant population of galaxies which habour a dormant BH at their centre \citep[see e.g.,][which shows that the general population of faint X-ray AGN peak at $z<1$]{2005A&A...441..417H}. As a consequence of their inactive nature, it is more difficult to be able to study the accretion processes, and immediate environment surrounding these quiescent BHs. However, TDEs provide us with a way to gain insight into the properties of these dormant BHs.

As AGN are found most of the time in a low-luminosity state \citep{2008ARA&A..46..475H}, it is possible that a flare arising from an inactive galaxy resulting from a major inflow of material onto the BH could ``reignite'' an AGN \citep[e.g.,][]{2006ApJS..163....1H}. In fact, it is thought that nearly 1-10\% of AGN detected in extragalactic X-ray surveys may actually be emission from a TDE \citep{2015MNRAS.452...69M}. Therefore, it is essential to better understand the differences between the emission arising from AGN detected in extragalactic surveys and that of a TDE such that we can distinguish between the emission from these events cleanly. 

Using the properties of the X-ray TDE and likely X-ray TDE sample derived by \citet{2016arXiv161102291A}\footnote{The events which \citet{2016arXiv161102291A} classified as X-ray TDEs include jetted events \textit{Swift J1644+57}, and \textit{Swift J2058+05} and non-jetted events \textit{ASASSN-14li}, and \textit{XMMSL1 J0740-85}. The likely X-ray TDE sample includes non-jetted TDEs: \textit{2MASX J0249}, \textit{XMM J152130.7+074916}, \textit{IGR J17361-4441}, \textit{NGC247}, \textit{OGLE16aaa}, \textit{PTF-10iya}, \textit{SDSSJ1201}, \textit{SDSSJ1311}, and \textit{SDSSJ1323}. This sample is used throughout the paper. }, we can compare how the peak X-ray emission from these events differs from that of AGN found at similar redshifts. In Figure \ref{lumredvshr} we have plotted the soft (0.3-2.0 keV) X-ray luminosity as a function of redshift for our TDE sample, and for a sample of AGN detected in the 4Ms Chandra Deep Field South \citep[CDFS:][]{2006A&A...451..457T, 2011ApJS..195...10X}, the 2Ms Chandra Deep Field North \citep[CDFN:][]{2003AJ....126..539A}, Chandra Multiwavelength Project \citep[ChaMPs:][]{2007ApJS..169..401K}, Chandra Large Area Synoptic X-ray Survey \citep[CLASXS:][]{2004AJ....128.1501Y} and the XMM-COSMOS survey \citep{2009A&A...497..635C}. The colours of each data-point represents the hardness ratio, HR=(H-S)/(H+S), of each source, where S is the number of counts in the soft 0.3-2.0 keV energy band, while H is the number of counts in the 2.0-10.0 keV energy band.

From Figure \ref{lumredvshr}, we find that the peak X-ray emission arising from a majority of our TDE sample is significantly brighter than the emission from an AGN found at the same redshift. In addition, we find that our non-jetted TDE sample indicated by the filled circles in our plot, have a peak X-ray luminosity that is significantly softer than that expected from an AGN. For our jetted TDE sample represented by the star symbols, and the emission from IGR J17361-444 which was originally detected in the hard X-ray energy band using INTEGRAL \citep{2014MNRAS.444...93D}, we have the opposite case. Here the emission arising from these sources is harder than that seen from AGN at the same redshift.

The fact that our current sample of X-ray TDEs are significantly brighter, and much softer than the general population of AGN found in deep X-ray surveys, indicates that these highly luminous prompt\footnote{Here prompt is defined as events which peak over a few weeks to months.} events would easily be spotted when one compares their properties to the general properties of the AGN population detected in the same survey. However, we should note that this separation in luminosity also indicates that current TDE studies are biased such that the brightest events are readily detected. \citet{2016arXiv161102291A} showed that the properties of these luminous TDEs implies that they are viscously slowed, indicating that there must be a significant population of low luminosity TDE events currently being missed or mistaken for other phenomenon. As a consequence, the luminosity distribution of X-ray TDEs seen in Figure \ref{lumredvshr} may not representative of the actual distribution of luminosities for these events, while low luminosity TDEs could potentially be contaminating the AGN population detected in extragalactic surveys as suggested by \citet{2015MNRAS.452...69M}.

\section{Variability and coherence of a detected flare}\label{va}

\subsection{Variability}

\begin{figure*}[t]
	\begin{center}
		\includegraphics[width=1.01\columnwidth]{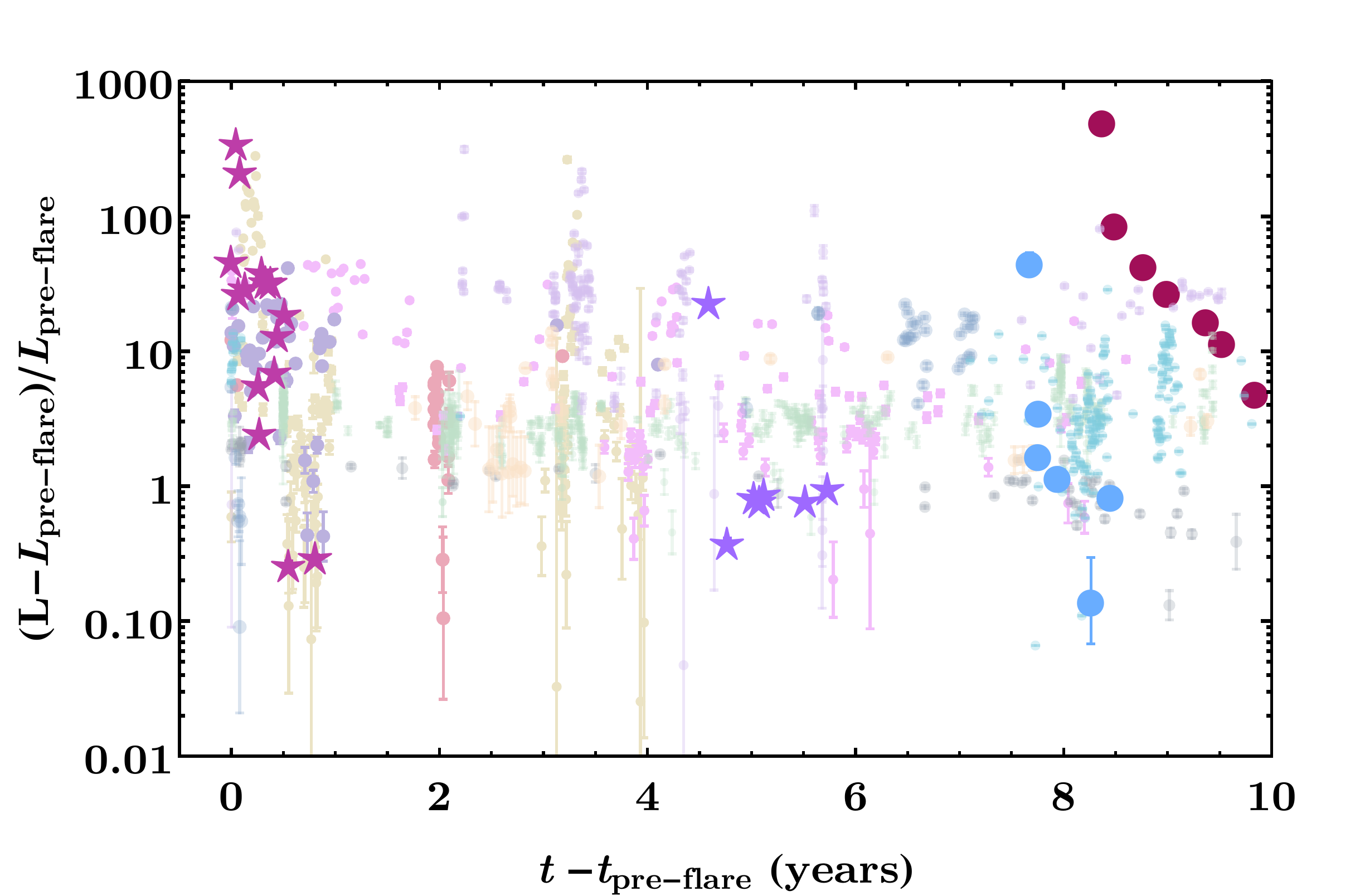}
		\includegraphics[width=0.99\columnwidth]{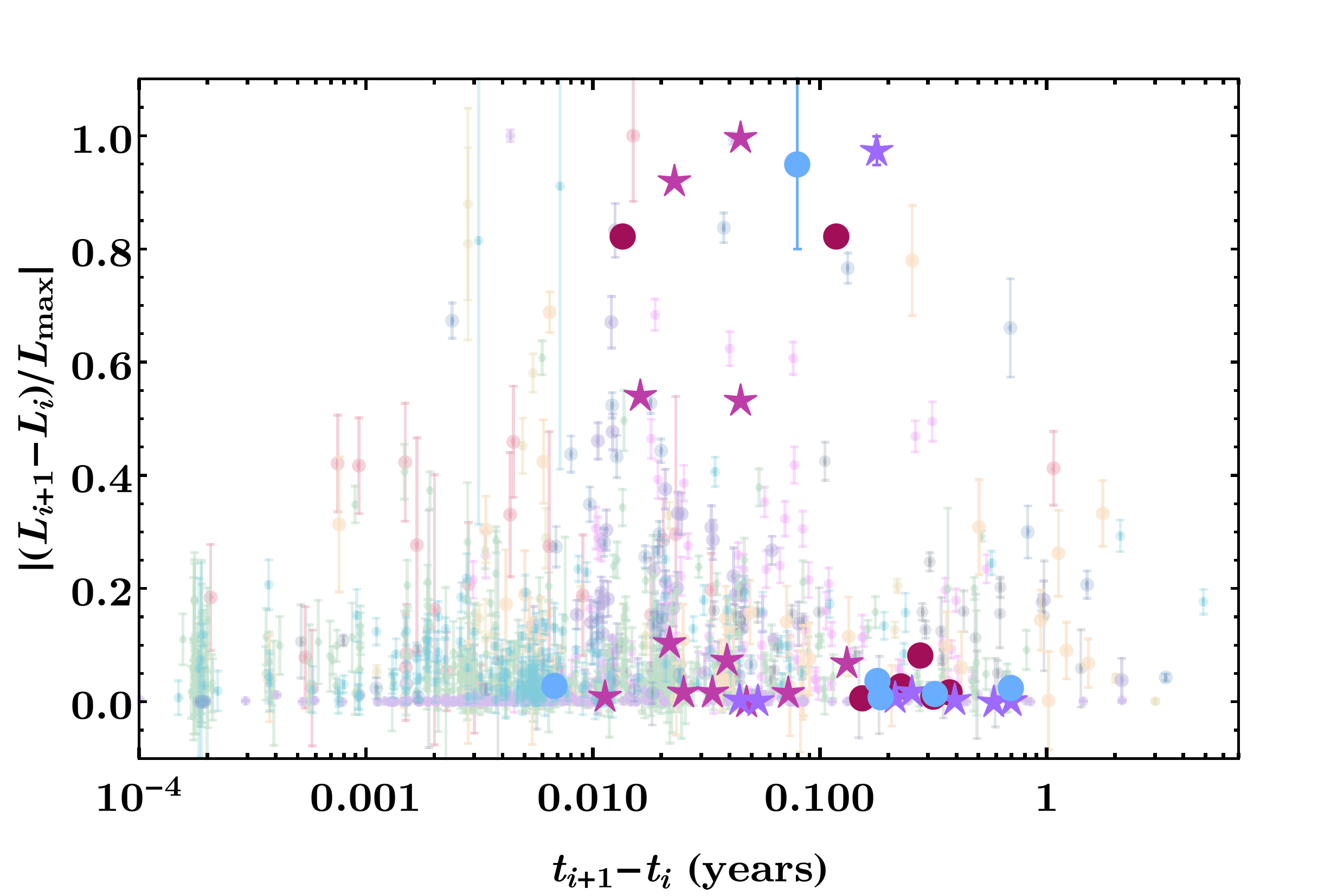}
				\includegraphics[width=0.7\textwidth]{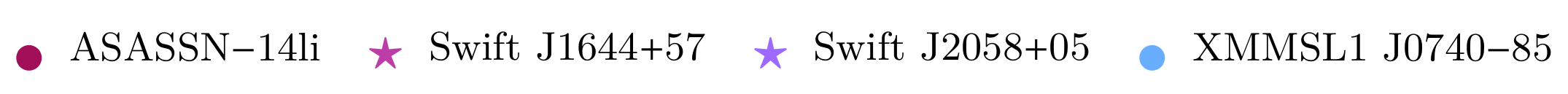}
		\includegraphics[width=0.5\textwidth]{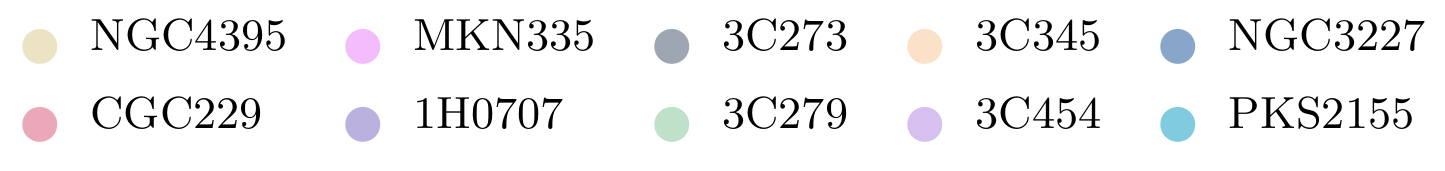}
		\caption{Metrics in which TDEs and AGN emission are indistinguishable. In the \textit{left panel} is the difference between the X-ray luminosity detected at or after peak, ((L-L$_{\rm pre-flare}$)/L$_{\rm pre-flare}$), compared to either a pre-flare upper-limit or pre-flare low state emission ($t-t_{\rm pre-flare}$) for our TDE and highly variable AGN sample. As the pre-flare upper-limit for TDEs \emph{ASASSN-14li} and \emph{XMMSL1~J0740-85} are from \emph{ROSAT} observations taken $\sim$22 years prior to detected emission from these events, we have scaled the ($t-t_{\rm pre-flare}$) for these events by 16 years so that they fit on our plot. In the \textit{right panel} is the magnitude relative to peak luminosity ($|(L_{i+1}-L_{i})/L_{\rm max}|$) in which the detected X-ray emission changes on small timescales ($t_{i+1}-t_{i}$). In both plots, TDEs plotted with a star ($\bigstar$) have been classified as jetted, while those plotted with a filled circle ($\bullet$) are classified as non-jetted in the literature. One can see that it is difficult to differentiate between the emission arising from an TDE and that from an AGN, assume that they had the same BH mass.\label{hvavstde}}
	\end{center}
\end{figure*}

\begin{figure}[t]
	\begin{center}
		\includegraphics[width=1.0\columnwidth]{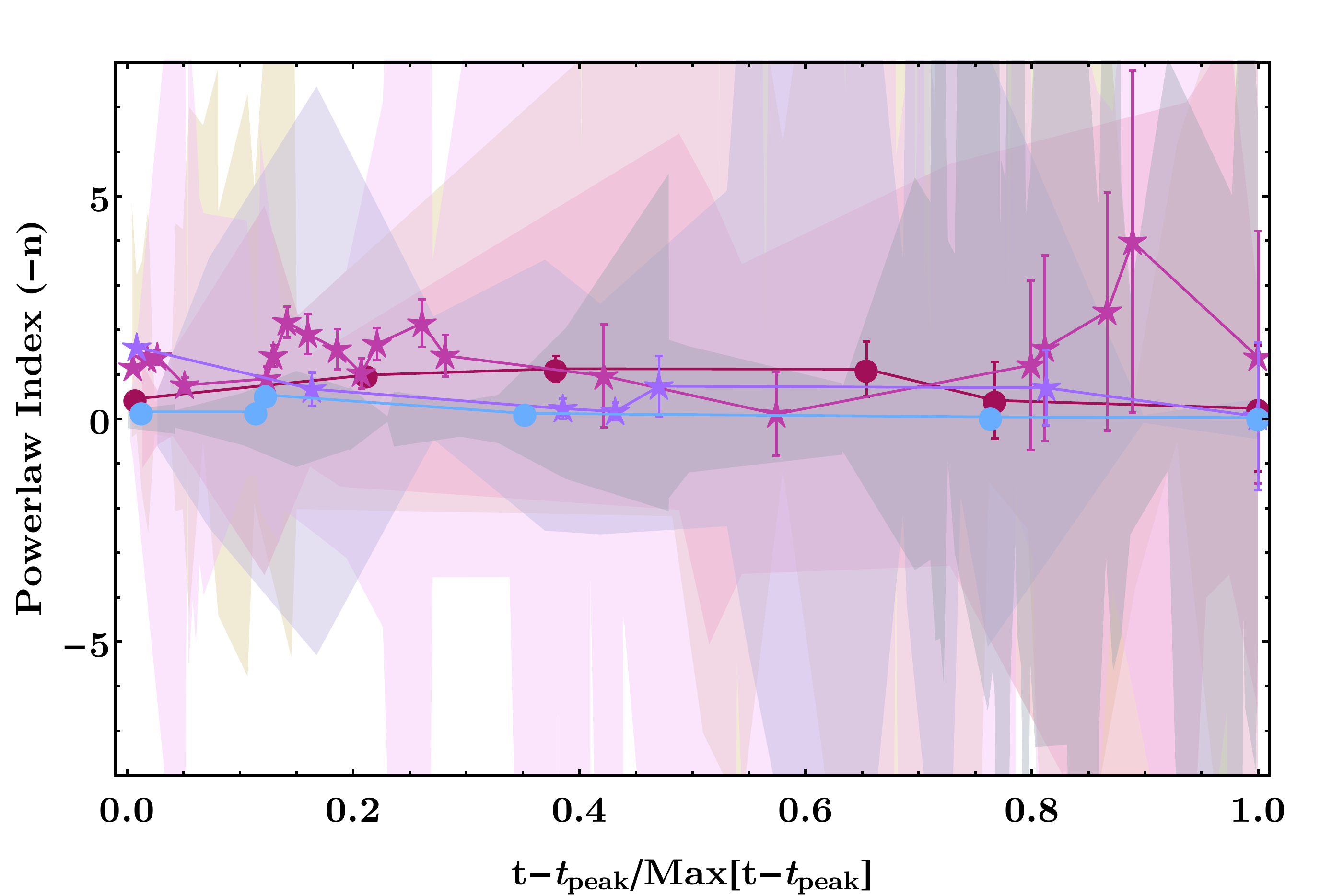}
			\includegraphics[width=1.0\columnwidth]{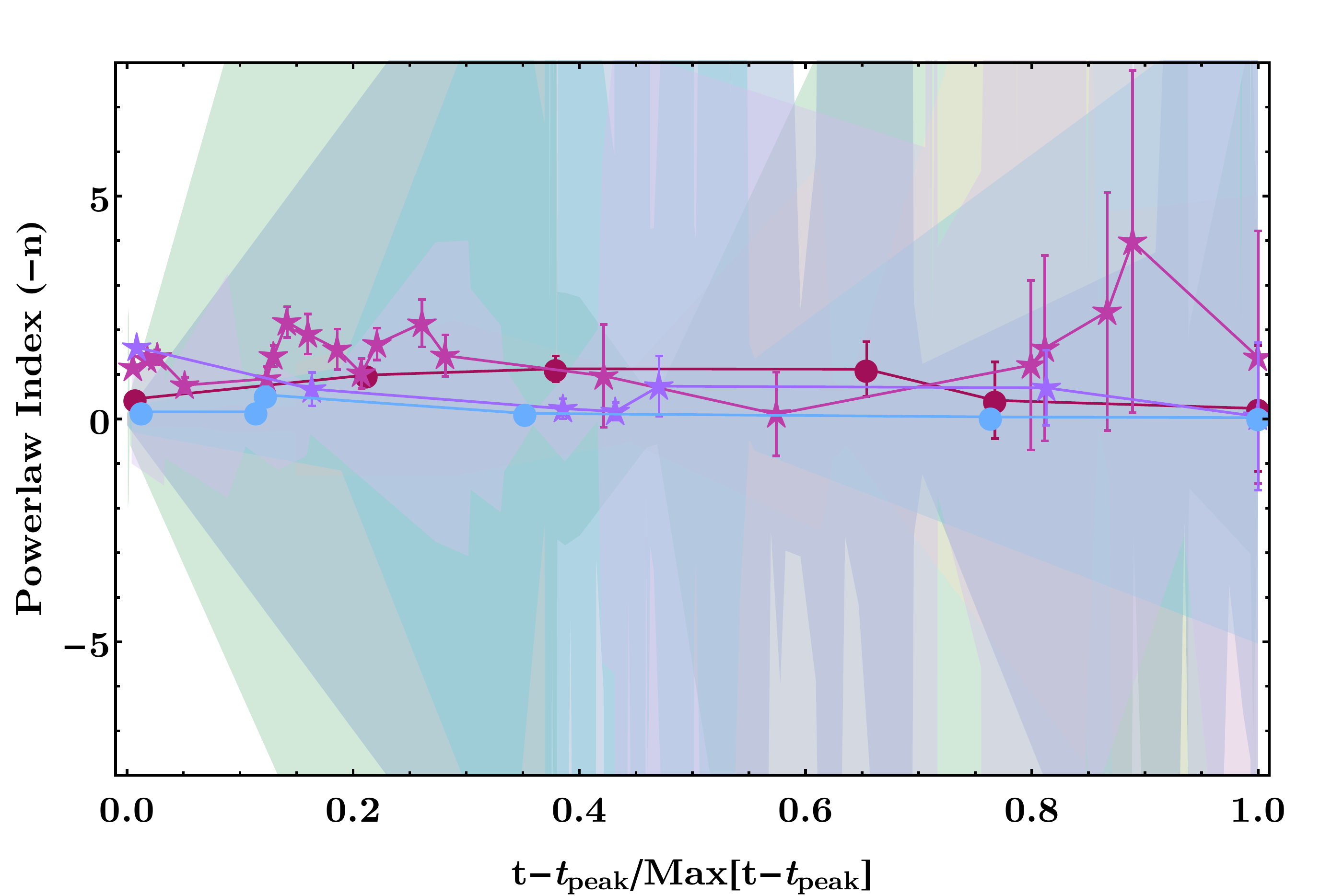}
		\includegraphics[width=0.9\columnwidth]{key.pdf}
		\includegraphics[width=0.7\columnwidth]{agnkey.pdf}
		\caption{A metric by which TDE emission is distinguishable from that of AGN. Here we have plotted the best fit powerlaw index ($-n$) and its uncertainty as time of peak goes to infinity, as a function of $t-t_{\rm peak}$. Due to the highly variable nature of our AGN sample we find significant variations in power-law index at $t_{\rm peak}\rightarrow\infty$. As such, we represent the full range of their derived power-law indexes and their uncertainties as different colour shaded bands.\label{hvavstde2}}
	\end{center}
\end{figure}

One of the defining characteristics of AGN is the detection of significant variation in their X-ray flux over a wide range of timescales \citep[][]{1976PASP...88..844H}. Studying this variation is key to better understanding the properties of AGN, and significant work has been done to characterise both the short term \citep[e.g.,][]{2012A&A...542A..83P} and long term \citep[e.g.,][]{2003AJ....126.1217D, 2011A&A...536A..84V, 2016arXiv161208547M} variability in these objects. It has been shown from their derived power spectral density functions that low luminosity AGN exhibit strong variability over short timescales of less than a day, while more massive AGN will show variability on timescales of years \citep[e.g.,][]{2006Natur.444..730M, 2005MNRAS.363..586U}. The origin of the short timescale variability is thought to arise from a hot corona
close to the central BH \citep[e.g.,][]{2004PThPS.155..170U, 2006Natur.444..730M}, while the variability observed on long timescales is thought to arise from disk instabilities \citep[e.g.,][]{1984ARA&A..22..471R,  1997ApJ...482L...9S}. In addition, e.g., \citet[][]{1997ApJ...476...70N} have shown that there is an inverse correlation between the amplitude of the short timescale variability of AGN and their luminosity. However, with the advent of multi-decade monitoring campagins of various AGN, it has become apparent that the amplitudes associated with their long term variability do not following this inverse relationship and instead are comparable between sources, indicating that these long timescale variations are set by the mass of AGN's BH \citep{2001ApJ...547..684M}.

Until relatively recently, it has been difficult to characterise with sufficient cadence, the long term X-ray light curves of X-ray TDE candidates to search for variability in their X-ray emission. As a consequence, our understanding of the differences and/or similarities of the emission from a flare detected from an AGN, and that detected from a TDE is inadequate at best. 

To shed light on this, we use the four X-ray TDEs listed in \citet{2016arXiv161102291A}, to characterise how the emission from these candidates differs from that of a sample of highly variable AGN that includes NLS1, BLS1, blazars and optically violent variable QSO sources. These highly variable AGN were chosen for this study as these source represent the most extreme variations seen in the emission arising from an AGN. In addition, they are known to produce flare-like signatures similar to that of TDEs \citep[e.g.,][]{2015A&A...581A..17C}. The sample of AGN we choose for this study include the four AGN selected by \citet{2015A&A...581A..17C} for their study based on large intrinsic variations in their X-ray emission as detected using \emph{Swift}. This includes NLS1 AGNs \textit{Mkn 335} \citep[e.g.,][]{2007ApJ...668L.111G} and \textit{1H 0707-495} \citep[e.g.,][]{2002MNRAS.329L...1B}; one of the least luminous, lowest mass and nearest Seyfert 1 Galaxy \textit{NGC4395} \citep{2003ApJ...588L..13F} which has also been classified as a BLS1/S1.9 \citep{2009MNRAS.394.2141N}, and \textit{CGCG 229-10} also called \textit{Zw 299-015} which is a Type I Seyfert Galaxy that exhibits broad-line emission \citep[e.g.,][]{2011ApJ...732..121B}. In addition to these four we also included broad-line Seyfert 1.5 AGN \textit{NGC 3227} \citep[see e.g.,][and references therein]{2009ApJ...691..922M}, high frequency BL Lac blazar \textit{PKS 2155-304} \citep[e.g.,][]{1999ApJ...527..719Z, 2005ApJ...629..686Z}, optically violent variable blazars \textit{3c279} \citep[e.g.,][]{1996ApJ...461..698H, 2008ApJ...689...79C} and \textit{3c345} \citep[e.g.,][]{1990A&A...227L..25K}, and quasars \textit{3c454.3} \citep[e.g.,][]{2006A&A...445L...1F, 2006A&A...456..911G} and \textit{3c273} \citep[e.g.,][]{1992A&A...254...96K, 2017ApJS..229...21X}. These sources are optimal for our analysis as they have a sufficiently large number of observations such that the emission from these events has a temporal coverage similar to those seen for our TDE sample.

To perform this analysis, we use the X-ray products (counts and light curves) that were extracted for the X-ray TDE events defined in \citet{2016arXiv161102291A}. For our AGN sample, we downloaded all available \emph{Swift} XRT observations for these objects and reprocessed the level one data from these observations following the standard procedure suggested in the \emph{Swift} XRT Data reduction guide\footnote{\url{http://swift.gsfc.nasa.gov/analysis/xrt_swguide_v1_2.pdf}}. Following the procedure defined in Section 2.2 of \citet{2016arXiv161102291A}, we extracted counts in both a soft (0.3-2.0 keV) and hard (2.0-10.0 keV) energy bands from each observation, and derive both the soft X-ray light curves and HRs for each of our highly variable AGN.

To quantify how luminous flares of X-ray TDEs are compared to those seen in AGN, we have plotted in Figure \ref{hvavstde} (left) the difference between the X-ray luminosity detected at and after the peak emission, relative to the luminosity of either a pre-flare upper-limit (for the TDEs) or a low state pre-flare detection (for the AGN) chosen immediately preceding the first X-ray detection of the flare ((L-L$_{\rm pre-flare}$)/L$_{\rm pre-flare}$). This is plotted against the difference in the time of a detection and the time of the pre-flare emission ($t-t_{\rm pre-flare}$). Due to the fact that the pre-flare upper-limit for TDEs \emph{ASASSN-14li} and \emph{XMMSL1~J0740-85} are from \emph{ROSAT} observations taken $\sim$22 years prior to detected emission from these events, we have scaled the ($t-t_{\rm pre-flare}$) for these events by 16 years so that they fit on our plot. 

From Figure \ref{hvavstde} (left) we can see that both TDEs and highly variable AGN can produce flare-like emission that increases in luminosity by one to three orders of magnitude relative to their detected pre-flare emission. This makes it difficult to differentiate between TDEs and highly variable AGN based on the detection of a many order increase in X-ray luminosity from a previously quiescent X-ray source.  Unlike TDEs, whose emission decays over timescales greater than a year, the flare emission from an AGN decays over a wide range of timescales, with some of the flares disappearing on timescales of a few months, while others take many years to decay back to pre-flare levels. As such focusing on how long a detected flare takes to decays is not sufficient for us to be able to differentiate between these two sources. 

A robust way of distinguishing AGN from TDEs is waiting for recurring flare-like emission which would rule out a flare as a TDE. However, one can see from Figure \ref{hvavstde} (left), even though AGN can produce multiple flare-like events over the same timescales that TDEs decay, AGN can also produce flare like emission over timescales that are significantly longer than the lifetimes of TDEs. Due to our uncertainty in the duty cycle of AGN, relying solely on the detection of a recurring flare emission from the position of a TDE is not the most optimal way of ruling out these events as AGN. 

The large variation in the behaviour seen in the X-ray light-curves of our AGN sample indicates that it is very difficult to be able to differentiate between these different objects using only their soft X-ray light curves. The fact that both TDEs and highly variable AGN produced flares of a similar order of magnitude, and decay timescales, has important implications for current and future X-ray studies that search for TDE candidates. Unless one has multi-year observations of a source (or preferably multi-decade) for which one can detect multiple flares, searching for these events by relying solely on detecting flare-like emission from a previous quiescent source will lead to a significant number of spurious detections. As a consequence, one requires other approaches to be able to qualify based on only the detected X-ray emission from a source whether the flare event is a TDE or not.

Another potential avenue in which one could differentiate emission from an AGN and that from a TDE is by looking at the magnitude at which the detected X-ray emission changes between different observations (i.e., as a function of timescale) during a flare for both TDEs and AGN. In Figure \ref{hvavstde} (right) we have plotted $|(L_{i+1}-L_{i})/L_{\rm max}|$ as a function of $t_{i+1}-t_{i}$, where $L_{i+1}-L_{i}$ and $t_{i+1}-t_{i}$ is the change in luminosity and time between consecutive observations $i$ and $i+1$. As we are interested in the magnitude of these luminosity changes, we have taken the absolute magnitude $L_{i+1}-L_{i}$ and normalised this value by the maximum luminosity detected from an observed flare. 

One can see that between consecutive observations, both AGN and TDEs can produce similar variations in the luminosity. There is a slight hint that TDEs possibly exhibit large changes in luminosities over longer timescales than AGN, however due to the relatively under-sampled nature of the X-ray light curve of TDEs compared to those of AGN, this might be an effect of systematics rather than an inherent property of these events. 

We find that the changes in luminosity experienced by AGN between consecutive observations cover a wider range of luminosities than those of TDEs. Here the majority of the emission from an AGN will exhibit changes between $(0.0-0.5)L_{\rm max}$, while the variations in luminosity from TDEs seems to be approximately bimodal where these changes fall into either changes that range between $(0.0-0.15)L_{\rm max}$ or $(0.8-1.0)L_{\rm max}$. This hint of bimodally might be a consequence of the sparser sampling of X-ray TDE lightcurves, however if true this indicates that AGN tend to show significantly more variability in their emission over all timescales, while even though TDEs will exhibit a dramatic increase in their X-ray emission during the initial flare, for the majority of their emission the variability between consecutive data points is significantly less, hinting towards the possibility that TDEs tend to have a coherent decay than AGN.

\begin{figure*}[!ht]
	\begin{center}
		\includegraphics[width=1.0\columnwidth]{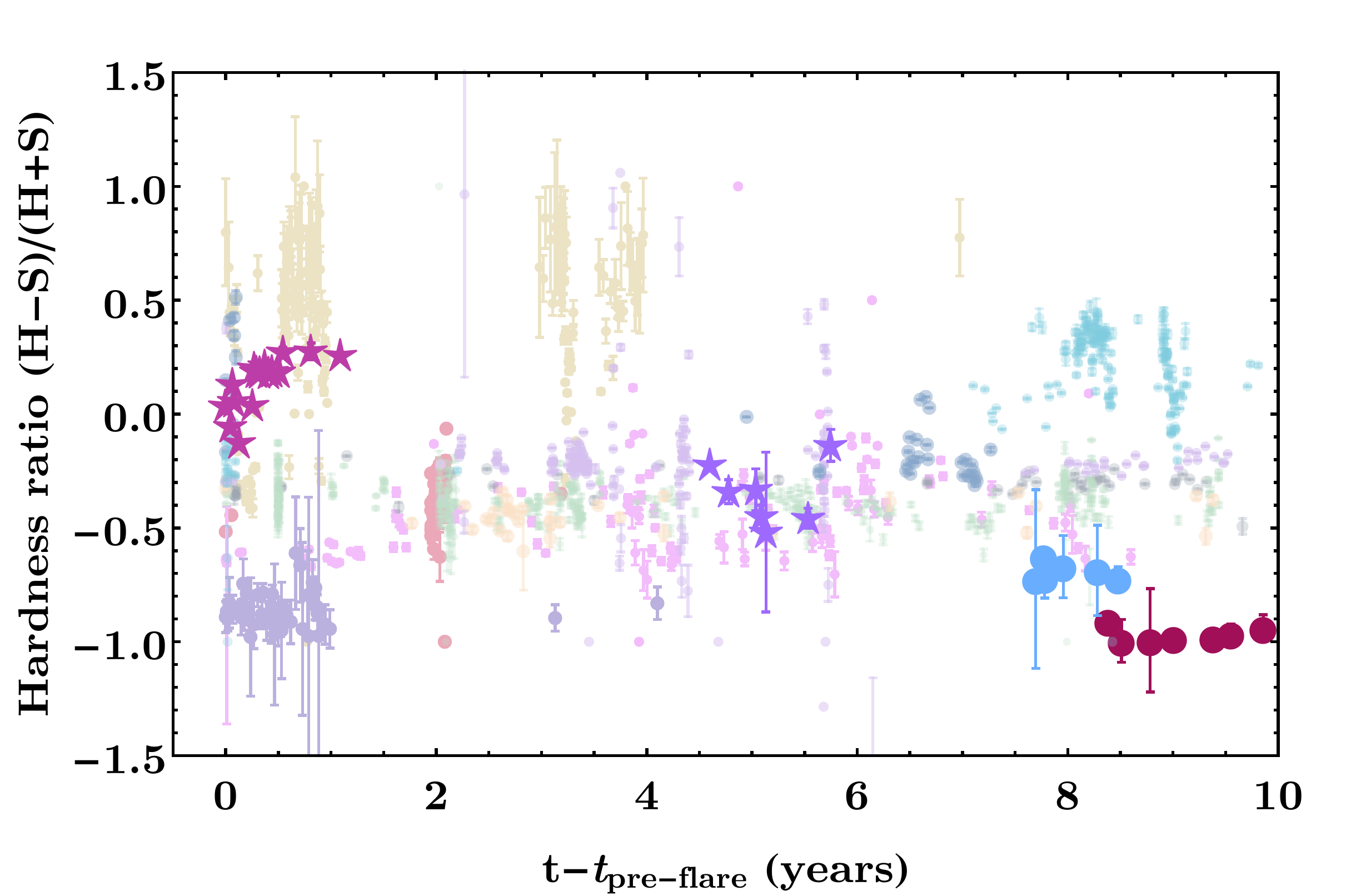}
		\includegraphics[width=1.0\columnwidth]{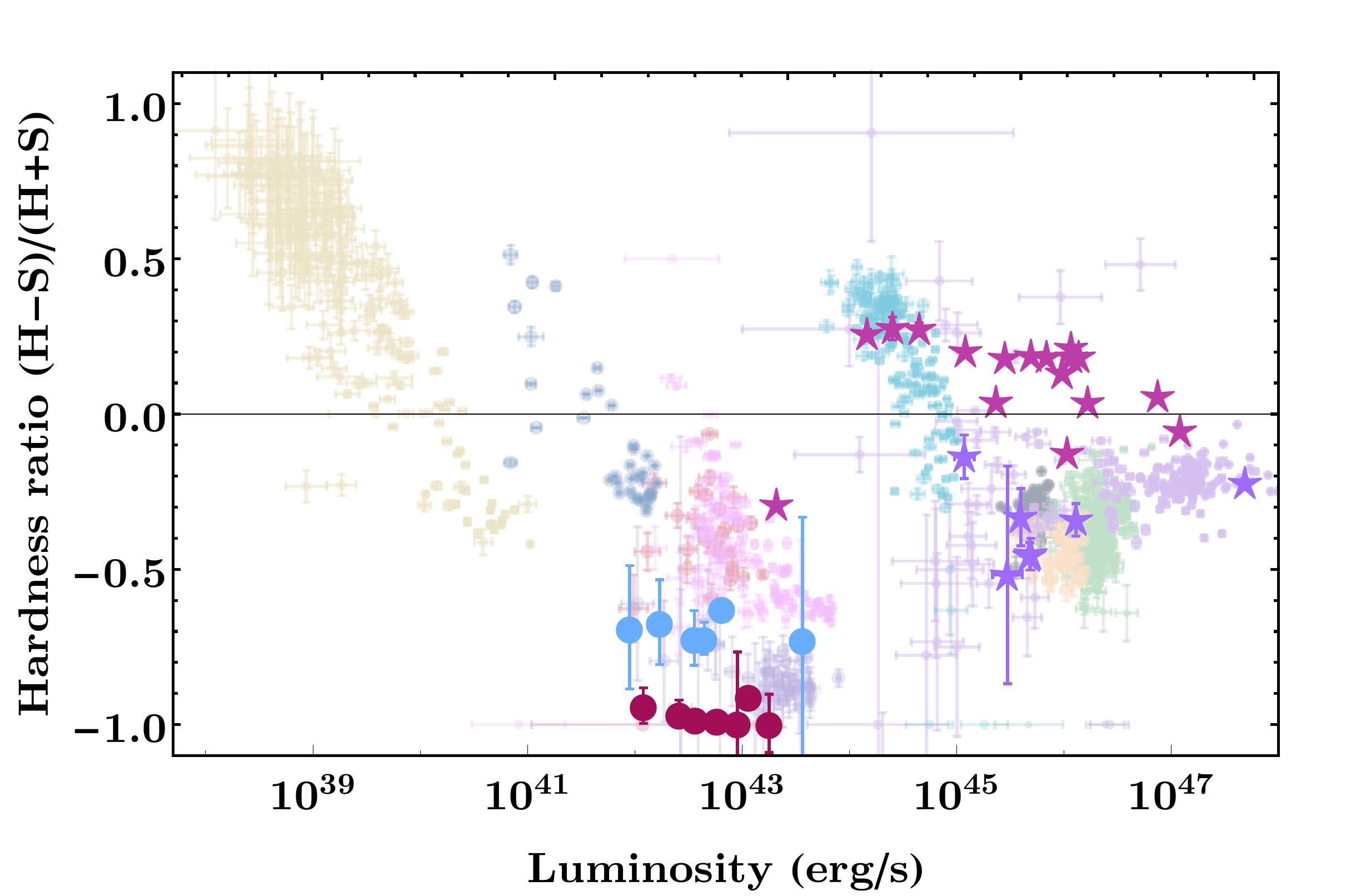}
				\includegraphics[width=0.7\textwidth]{key.pdf}
\includegraphics[width=0.5\textwidth]{agnkey.pdf}
		\caption{Hardness ratio defined as (H-S)/(H+S), where H is the counts in the equivalent 2.0-10.0 keV energy band and S is the counts in the equivalent 0.3-2.0 keV energy band, as a function of \textit{left:} $t-t_{\rm pre-flare}$ (years) and \textit{right:} soft (0.3-2.0 keV) X-ray luminosity (erg cm$^{-1}$) for our TDE and highly variable AGN sample. Here the colours and symbols are defined in Figure \ref{hvavstde}, while in the \textit{left panel} we have also scaled $t-t_{\rm pre-flare}$ for \textit{ASASSN-14li} and \textit{XMMSL1 J0740-85} as what was done in Figure \ref{hvavstde} left panel. \label{colors}}
	\end{center}
\end{figure*}

\subsection{Coherence}\label{coh}

To investigate how coherent in time the emission from a TDE is compared to that of an AGN, we determined the best fit power-law index ($-n$) for both the TDE and AGN sample as the time of peak goes to infinity assuming $L\propto (t-t_{\rm peak})^{-n}$, similar to what was done in \citet{2016arXiv161102291A}. We have plotted the  $n$ as a function of $t-t_{\rm peak}$ in Figure \ref{hvavstde2}, where $t_{ \rm peak}$ is the time in which we detect the peak luminosity. Due to the fact that our sources decay over different timescales, we have normalised $t-t_{\rm peak}$ by the maximum value of $t-t_{\rm peak}$ for each of our sources such that we can compare the differences in behaviour. When completing this exercise we found that the derived power-law indexes for AGN varied quite significantly over short time scales, ranging between a power-law index of $n\sim-10$ and $n\sim+15$. Due to the highly variable nature of the AGN emission and thus the large variations in the power-law index, we represent the full range of power-law indexes and their one sigma uncertainties as a different coloured shaded band, rather than solid symbols as we have done for the TDEs. 

From Figure \ref{hvavstde2} the differences in how a flare arising from a TDEs decays compared to that of a highly variable AGN is quite striking. While the emission from a highly variable AGN varies wildly during its decay, the emission from a TDE is much more coherently, exhibiting approximately a monotonic decay during its flare. As a flare from a TDE decays, it is best fit with a powerlaw index that ranges between $n=0$ and $n\sim-2$. However, for a majority of its decay, the powerlaw index changes little, especially for the two non-jetted TDEs \emph{ASASSN-14li} and \emph{XMMSL1 J0740-85}. The jetted TDEs \emph{Swift J1644+57} and \emph{Swift J2058+05} do show  significantly more variability in their best fit powerlaw indexes compared to that of the non-jetted events, however this variation still falls with $n=0$ and $n\sim-2$, and is significantly less dramatic compared to the AGN sample. 

Even though our results suggest that TDEs decay more coherently than AGN, it is possible that our small sample of highly variable AGN with well defined light curves is not fully representative of the emission behaviour of highly variable AGN. As such, we could potentially be missing the fact that AGN can produce flares which follow a coherent decay similar to that of a TDE. To estimate the probability that an AGN could produce TDE-like flare behaviour, we randomly extract from each of the AGN light curves a set of five data points (similar to that seen for XMMSL1 J040-85), and fit them with a power-law decay law model as $t_{\rm peak}\rightarrow\infty$ to see if we can reproduce the coherent decay of a TDE seen in Figure \ref{hvavstde2}. We simulated and fit 10,000 light curve realisations and found that $<$1\% of these produced the coherent decay behaviour seen from our TDE sample. To make this more concrete, we calculate the probability that an AGN is responsible for the coherent flare we detect (i.e., what is the probability that our TDE sample arises from a highly variable AGN)\footnote{Here we are explicitly calculating the following: \begin{equation*}
P(\rm AGN|coherent)=\frac{P(\rm coherent| AGN)\,P(\rm AGN)}{P(\rm coherent|AGN)\,P(\rm AGN)+P(\rm coherent|TDE)\,P(\rm TDE)}.
	\end{equation*} Using the X-ray TDE plus the not a TDE sample from \citet{2016arXiv161102291A} as a guide, we assume that $P(\rm TDE)=20\%$, while based on our simulations we expect that 1\% of AGN flares are coherent ($P(\rm coherent|AGN)=1\%$).}. From this calculation, we determine that $\lesssim4\%$ of coherent flare emission we detect would arise from an AGN, suggesting that the emission from a TDE will decay more coherently than that of a highly variable AGN in the soft X-ray band. As such it is quite likely when one detects coherent emission from a flare at the centre of its host galaxy, that its origin is that of a TDE.

\section{Spectral hardness evolution}\label{hr}

A simple way to probe the properties of X-ray sources is to derive a hardness ratio (HR), which is especially useful for sources in which we are photon limited either due the to faintness of the X-ray emission from the source, or due to the limited exposure time of an observation. Depending on whether a source is hard or soft, this can inform us of either different accretion processes occuring in different objects, as well as how dust/gas-obscured a source is. If variations in the hardness ratio are detected, this can also indicate changes in the accretion model \citep[e.g.,][]{2008NewAR..51..733N}. 

The hardness of the X-ray emission arising from an AGN over various redshifts have been studied extensively in a deep extragalactic X-ray survey using \emph{Chandra} \citep[e.g.,][]{2009ApJ...696..891H, 2011ApJS..195...10X, 2016ApJ...830..100M} and \emph{XMM-Newton} \citep[e.g.,][]{2009A&A...497..635C, 2015A&A...573A.137L}. It has been shown that AGN with extended optical counterparts tend to exhibit harder X-ray emission due to the presence of significant gas absorbing softer photons, while AGN without a resolved counterpart tend to be softer \citep{2009ApJ...696..891H}. Due to the limited number of X-ray TDEs detected, it was not until recently have we been able to better characterise the spectral hardness of these events as a class \citep{2016arXiv161102291A}. However, unlike AGN which are much harder the more absorbed they are, \citet{2016arXiv161102291A} found that the emission detected from an X-ray TDE is quite soft in nature, even though they are highly absorbed.

Nevertheless, comparison of the spectral hardness (i.e., HR) of AGN and X-ray TDEs has been limited, and current studies have focused on how the emission from an individual TDEs events compare to the HRs of AGN derived from extragalactic X-ray surveys \citep[e.g.,][]{2017arXiv170200792L}. As such, it has not yet been quantified systematically how the spectral hardness of an AGN or a TDE flare changes as a function of time or luminosity. Using our X-ray TDE and highly variable AGN samples, we can now quantify this relationship.

In Figure \ref{colors} (left), we have plotted HR as a function of $t-t_{\rm pre-flare}$, while in Figure \ref{colors} (right) we have plotted HR as a function of X-ray luminosity in the 0.3--2.0 keV energy band. From these plots, one can see that a flare from an AGN tends to exhibit significant variation in HR both as a function of time and brightness (i.e., luminosity). At the peak of an AGN flare, their emission tends to be quite soft in nature, which could indicate that the flare is ionising the surrounding material making it transparent to soft X-ray photons. However, as the flare decays, and thus becomes fainter, the emission becomes harder. 

For TDEs though, the story is quite different. For non-jetted TDEs, the emission from these objects is very soft in nature. However, unlike AGN in which the emission becomes harder as it decays, we find that the emission from a non-jetted TDE flare shows little variation, and does not change significantly from its spectral hardness measured at peak. At peak, we find that jetted TDEs are harder than their non-jetted counterparts, or the peak emission seen from some of our AGN sample. We also see some variation in the HR of a jetted TDE flare as it decays, with its emission becoming harder with time (or as it becomes fainter). However, the HR variation seen for jetted TDEs is much smaller than what we see from our AGN sample. We also find that AGN can produce emission that can range from completely thermal (HR$=-1$) to complete non-thermal (HR$=+1$) and anywhere in between, while TDEs tend to be much softer and have a HR$\lesssim0.3$. 

The fact that an X-ray TDEs shows very little to no colour evolution compared AGN flares highlights how the interplay between the different accretion processes in these objects, and properties of their environment, can leave a significant imprint on their corresponding X-ray properties. Compared to AGN whose flares result from disk instabilities and optically thin material located close to the BH, the approximately steady-state accretion flow of material experienced by the BH during a TDE could produce an inherently more coherent spectral decay for these events. As variations in HR can also arise from dust obscuration, the fact that TDEs do not show significant colour evolution while exhibiting coherent decay (see Section \ref{coh}) implies that the BHs responsible for TDEs have a significantly simpler accretion structure surrounding them compared to AGN.  

In addition to their accretion disk, AGN are known to be surrounded by a significant amount of material (dusty torus), that reprocesses emission from its central engine and inner disk \citep[e.g.,][]{1988ApJ...329..702K, 2001ApJ...555..663S, 2009ApJ...705..298M}. As this torus is known to be quite clumpy \citep[e.g.,][]{2007A&A...474..837T, 2009ApJ...697..182T}, this can lead to significant variations in the emission arising from an AGN. However, for TDEs, \citet{2014ApJ...783...23G} suggested that this additional layer of material may either not form at all (in the case of viscously delayed events), or may form in a very compact way such that the central regions of the accretion disk are visible through material that is close to the Compton thick limit. If present, this layer is expected to be short lived and thus would have little affect on the emission of the event. As a consequence, one does not see these significant variations in both the observed flux and HR as seen in AGN.

Additionally, \citet{2004AJ....127.1799G} showed that these highly variable AGN have an Eddington luminosity ratio ($L/L_{\rm Edd}$) close to one, significantly higher those seen from the wider AGN population. Sources with large $L/L_{\rm Edd}$ tend to have strong outflow/winds \citep{1998A&A...331L...1S} which can easily move material directly surrounding the central engine in and out of our line of sight. As a consequence, the nature of these sources could also explain, at least to some extent, why we see such significant variation in HR and flux (see Section \ref{va}) for these objects. The fact that we still see some variation in HR and flux for our jetted events, but not at the same levels seen from our variable AGN sample suggests that either the column densities around these objects is significantly lower than those of AGN (see Section \ref{nhpwlcdfs} for more details), or these events have a  $L/L_{\rm Edd}$ less than that of our AGN sample such that it cannot produce a strong wind which would move material around. However, to form a jet, \citet{2012ApJ...760..103D} suggested that one requires super-Eddington accretion rates, which would imply that our jetted TDE sample have $L/L_{\rm Edd}>1$. As a consequence, unlike our AGN sample, the highly collimated emission formed during a jetted TDE is significantly less affected by its surrounding material as it is produced far from the BH.

\begin{figure}[t]
	\begin{center}
		\includegraphics[width=0.49\columnwidth]{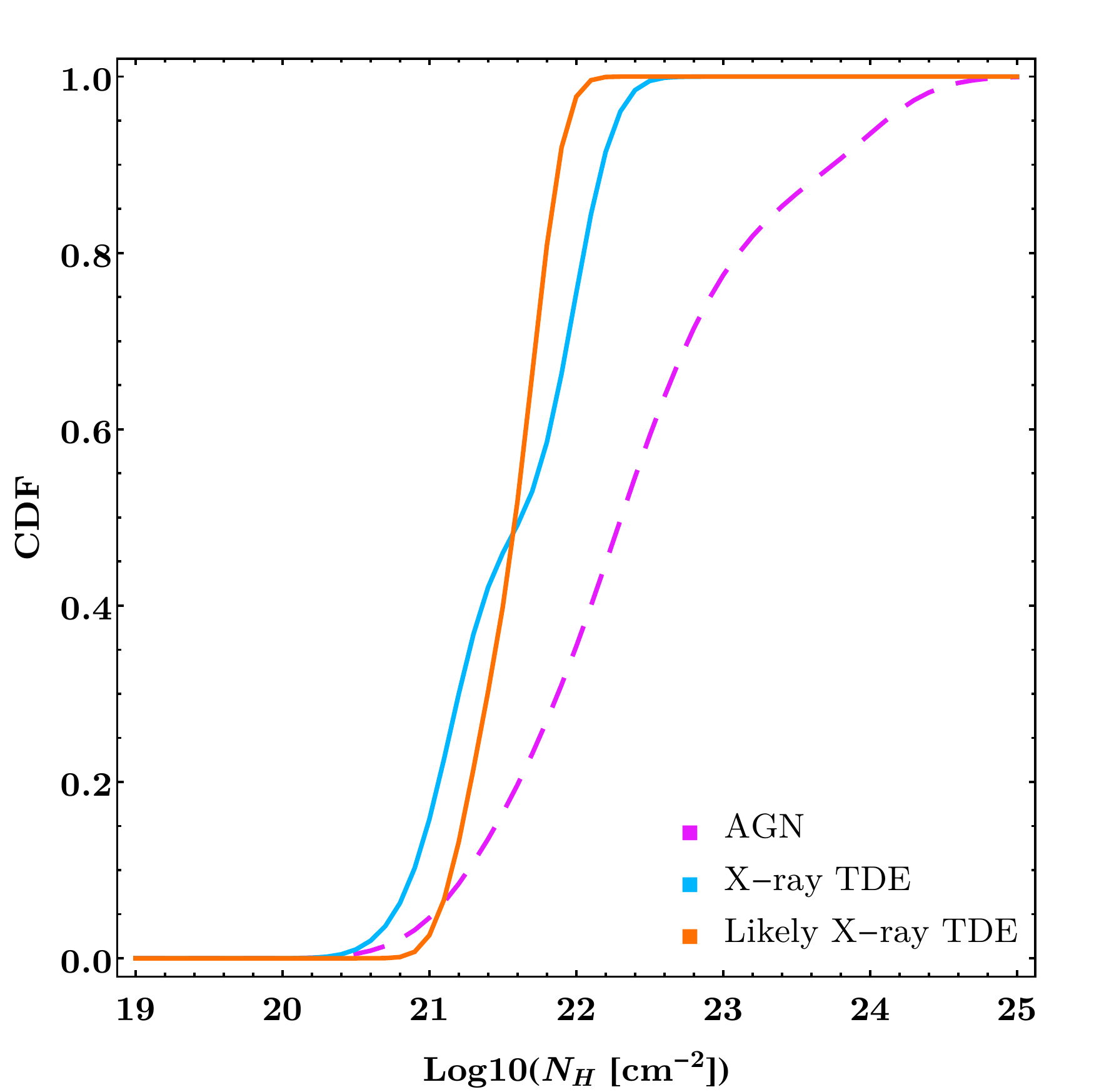}
		\includegraphics[width=0.49\columnwidth]{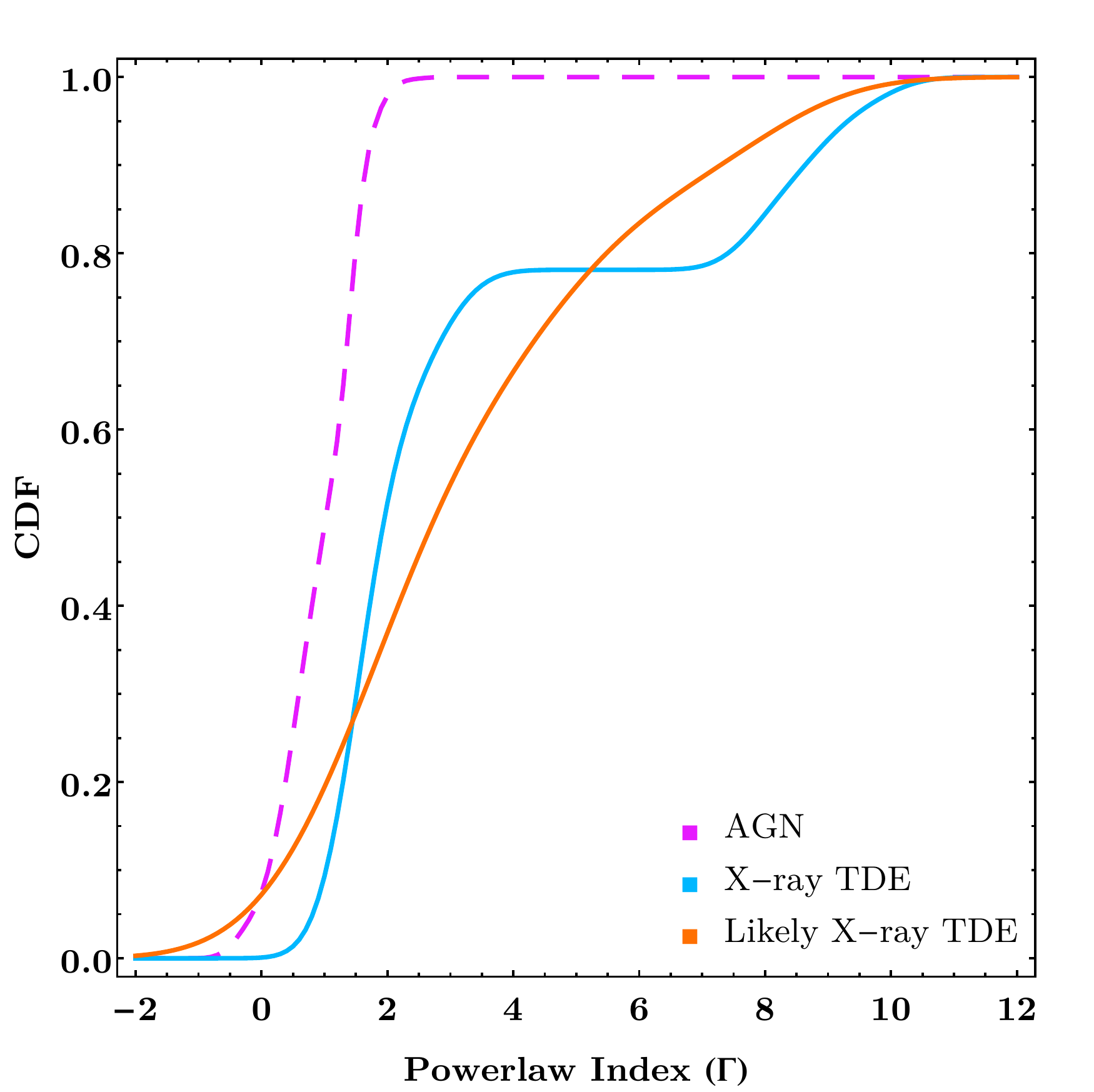}
		\caption{The cumulative distribution functions of \textit{left:} column density ($N_{H} [\rm cm^{-3}]$) and \textit{right:} powerlaw index ($\Gamma$) for X-ray TDE classifed as a X-ray TDE or likely X-ray TDE by \citet{2016arXiv161102291A}, and AGN detected using the Chandra Deep Field South \citep{2006A&A...451..457T, 2011ApJS..195...10X} given by the cyan solid, orange solid and magenta dashed curves respectively. \label{CDFs}}
	\end{center}
\end{figure}

\section{Spectral properties}\label{nhpwlcdfs}

The X-ray properties of the AGN population up to a redshift of $z\sim5$ have been extensively studied using multiple deep extragalactic X-ray surveys \citep[e.g.,][and references therewithin]{2006A&A...451..457T, 2008ApJ...681..113T, 2015A&A...573A.137L, 2016ApJ...830..100M, 2016MNRAS.463..348V}. From these studies, it has been found that there are several populations of AGN which can be distinguished based on their column density ($N_{\rm H}$). These include unobscured AGN which have a $N_{\rm H}<10^{22}$ cm$^{-2}$, obscured AGN which have a $N_{\rm H}>10^{22}$ cm$^{-2}$ and Compton thick AGN that have $N_{\rm H}>10^{24}$ cm$^{-2}$. In the low-redshift ($z<1-2$) universe, the AGN population is dominated by unobscured and obscured AGN, and the X-ray emission from these objects is best fit with a power-law index $\Gamma \sim1.75$ \citep[e.g.,][]{2006A&A...451..457T, 2016ApJ...830..100M}. 

As we are interested in comparing the properties of our X-ray TDE sample to those of AGN, we derive the cumulative distribution functions (CDFs) of $N_{H}$ and $\Gamma$ for X-ray TDEs and AGN found at $z<2$ (Figure \ref{CDFs}). For our TDE sample we use the $N_{H}$ and $\Gamma$ derived by \citet{2016arXiv161102291A}, while we use the $N_{H}$ and $\Gamma$ derived from the Chandra Deep Field South \citep{2006A&A...451..457T, 2011ApJS..195...10X} for our AGN. 

From Figure \ref{CDFs}, one can see that X-ray TDEs and AGN have significantly different X-ray emission properties. To quantify this, we derive the Cohen d statistic, which allows us to determine the mean difference between these two populations. We find the $N_{H}$ and $\Gamma$ for TDEs and AGN differ by at least 1.2 standard deviations (i.e., a d-statistic of 1.2 and 2.0 for $N_{H}$ and $\Gamma$ respectively.) From these plots, we find that X-ray TDEs are significantly less absorbed compared to AGN found at similar redshifts, while their emission is significantly softer than the emission detected from the same AGN population. The relatively unobscured nature of TDEs compared to that of AGN could be responsible for the fact that we observe large variation in the derived $\Gamma$ when one models the X-ray emission of TDEs. However, it has been shown that even unabsorbed AGN that have a $N_{H}<10^{22}$ cm$^{-2}$ have a power-law X-ray spectrum consistent with $\Gamma \sim1.8$ \citep[e.g.,][]{2006A&A...451..457T}. This indicates that emission from an AGN is intrinsically harder than that of a TDE, while the difference in environments lead to TDEs having a significantly wider variety of $\Gamma$ when modelling their X-ray emission.

\section{Soft X-ray luminosity function of X-ray TDEs}

\begin{figure*}[t]
	\begin{center}
		\includegraphics[width=1.0\columnwidth]{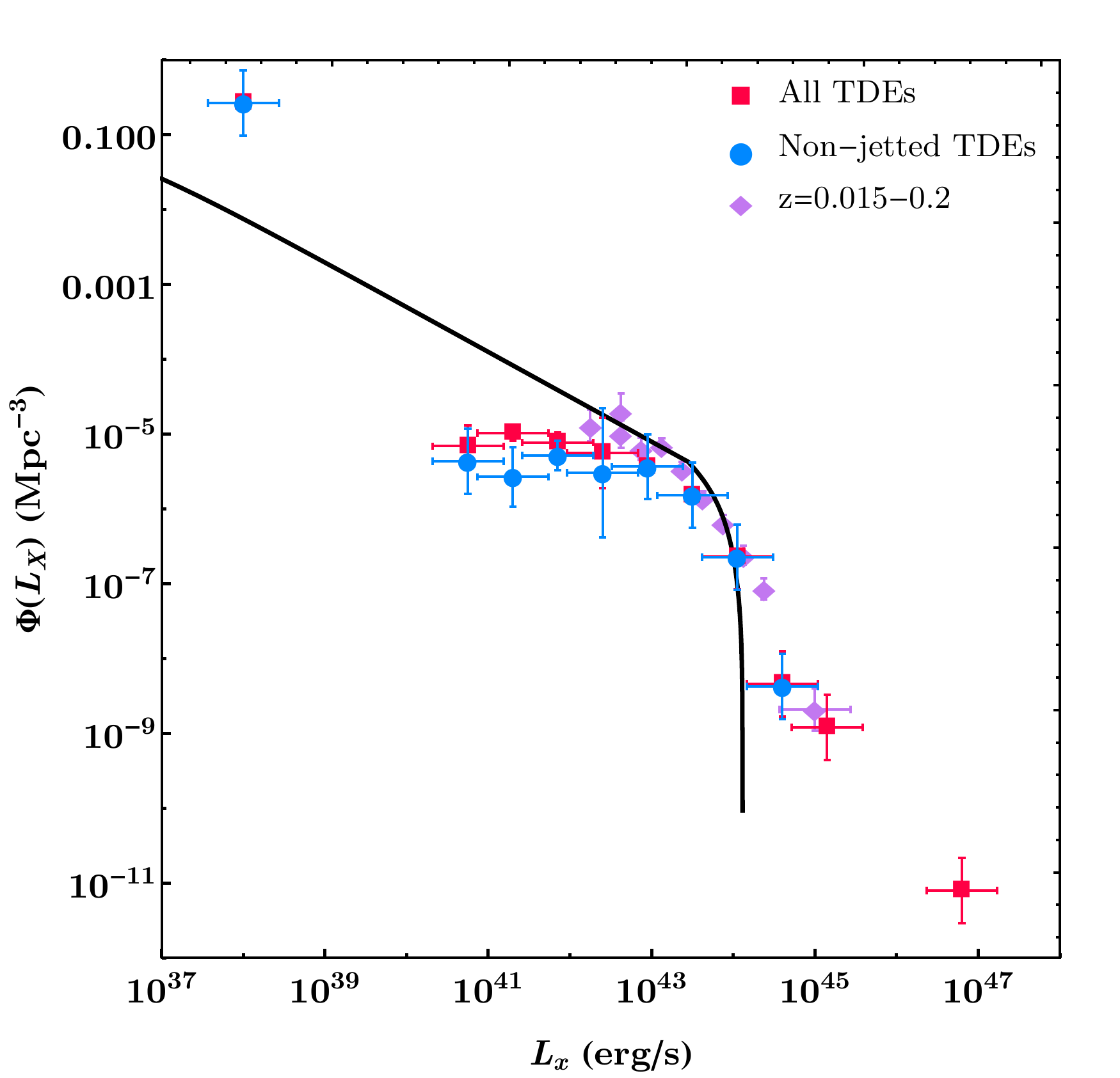}
		\includegraphics[width=1.0\columnwidth]{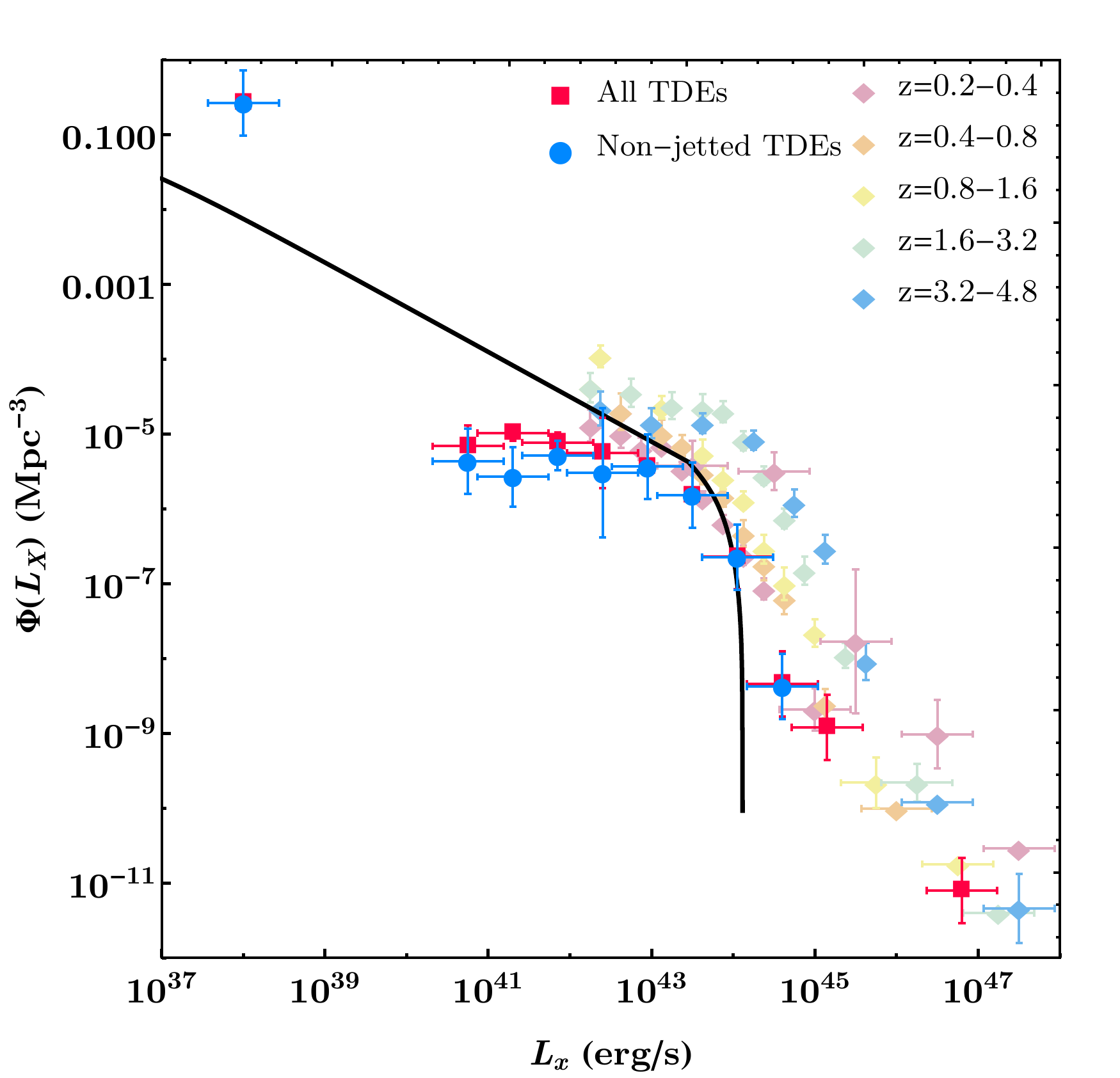}
		\caption{The soft X-ray luminosity function (LF) of X-ray TDEs derived using the emission from both jetted and non-jetted TDEs (red squares) and non-jetted events only (blue squares). We have also overlaid the soft X-ray LF for AGN for a redshift shell of 0.015--0.2 (\textit{left panel}) and for redshift shells greater than 0.2  (\textit{right panel}) as derived by \citet{2005A&A...441..417H}. These are plotted using diamonds and the different redshift shells are presented as various colours listed in the legend of each figure. Shown as the black curve is the theoretical X-ray LF for non-jetted TDEs derived by \cite{2006ApJ...652..120M}.  As current extragalactic X-ray surveys are flux limited, current studies have been only been able to constrain the AGN LF for $L_{X}>10^{42}$ erg s$^{-1}$. \label{XLF}}
	\end{center}
\end{figure*}

Deep X-ray observations using \emph{ROSAT}, \emph{Chandra} and \emph{XMM-Newton} \citep[see review by][and references therewithin]{2015A&ARv..23....1B} have provided us with the ability to resolve a large fraction of the extragalactic X-ray background. From these observations, and extensive follow-up campaigns, we now have a reasonably complete sample of both soft and hard X-ray selected AGN over a wide range of redshifts. Using this sample, a number of authors \citep[e.g.,][]{2001A&A...369...49M, 2003ApJ...598..886U, 2005AJ....129..578B, 2015MNRAS.451.1892A, 2015ApJ...804..104M} have been able to quantify the AGN luminosity function (LF) and its evolution as a function of redshift, providing a key observational constraint on the origin and accretion history of the SMBH that reside at the centre of most galaxies. 

It has been shown by e.g., \citet{2006ApJ...652..120M} and \citet{2013ApJ...777..133M}, that TDEs can power low-luminosity AGN. As a consequence, it is thought TDEs could contribute significantly to the LF of AGN derived in X-rays. \citet{2006ApJ...652..120M} and more recently \citet{2015MNRAS.452...69M}, theoretically estimated the contribution of non-jetted TDEs to the X-ray LF of AGN and found that regardless of the various uncertainties that arise from this calculation, TDEs can contribute significantly to the AGN LF for luminosities $<10^{44}$ erg s$^{-1}$. For luminosities $>10^{44}$ erg s$^{-1}$, the contribution of TDEs is negligible due to the fact that $\sim10^{44}$ erg s$^{-1}$ corresponds to the Eddington limit associated with the rate of accretion for BHs with masses $M_{\rm BH}\sim10^{6-7}M_{\odot}$. Above this mass, the tidal radius of a main sequence star lies with the Schwarzschild radius of the BH, and thus these stars would be swallowed whole rather than disrupted \citep[see][]{2012PhRvD..85b4037K, 2017arXiv170108162L}. 

Currently, most of our understanding related to how the number of TDEs changes with luminosity has come from theoretical studies like those of \citet{2006ApJ...652..120M}. However, it would be beneficial if we could constrain observationally the LF of TDEs, as this would provide us with a way to experimentally constrain the contribution of TDEs to the growth of BHs in the local universe, while providing a useful tool to predict the number of possible events one could detect in future X-ray surveys using instruments such as \emph{eROSITA}\footnote{See \url{http://www.mpe.mpg.de/455799/instrument}}. 

Recently, \citet{2016PASJ...68...58K} made the first attempt to characterise observationally the hard (4-10 keV) X-ray LF of TDEs using a flux limited sample of TDE events detected using 37 months of \emph{MAXI} data. From their study, they find TDEs do not contribute significantly to the hard X-ray LF of AGN for $z<1.5$, and thus their contribution to the BH growth at these redshifts is negligible.
Unfortunately, due to the limited number of TDEs that emit significantly in the hard X-ray energy band, \citet{2016PASJ...68...58K} likely miss a large number of TDE candidates that could be used to derive the X-ray LF. As X-ray TDEs are found predominantly in the soft X-ray band, using TDEs detected in this energy band would provide a more complete sample of events for us to use to determine the X-ray LF of TDEs. 

To derive the X-ray LF one requires a ``snap-shot'' of the TDE rate, which one can derive once a survey has observed a given area of the sky twice. Here the ``snap-shot'' rate is the number of transient events detected within the search volume of a survey above a certain flux limit\footnote{To see an example of how one derives a ``snap-shot'' rate, see Equation 5 of \citealt{2011MNRAS.417L..51V} who applied this type of calculation to derive the snap-shot rate of radio jets in TDEs.}. Some of the current X-ray searches for TDEs effectively behave like a ``snap-shot'' survey, for example \emph{ROSAT} which performed an all sky survey is a good example of this type of survey, and to some extent the \emph{Swift}-BAT also fulfils this requirement.  Due to the nature of how \textit{XMM-Newton} observes the sky during slew mode, it effectively takes snapshots of the X-ray sky, while it has covered approximately 85\% of the sky after correcting for overlap\footnote{See \url{https://www.cosmos.esa.int/web/xmm-newton/xmmsl2-ug} for more details}. Thus for our purposes, we assume that it effectively behaves like a ``snap-shot'' survey. As \emph{Chandra} does not offer an all-sky survey mode due to the nature of this instrument, while most TDE sources are followed up using \emph{Swift} or \emph{XMM-Newton}, we do not consider \emph{Chandra} as a ``snap-shot'' survey and thus we overlook its contribution to the LF for this study. 

Due to the different selection efficiency associated with the various surveys, one needs to consider the possibility that we are missing a large fraction of events. It was shown in Figure 5 of \citet{2016arXiv161102291A}, that nearly all of their X-ray TDE and likely X-ray TDE candidates would be detected using the \emph{ROSAT} all sky survey, while nearly half of these events would be seen using the \textit{Swift}-BAT. The \textit{XMM} slew survey has a similar flux limit as \emph{ROSAT} \citep{2016A&A...588A.103B}. Unfortunately, due to the nature of current X-ray instruments and triggering criteria associated with following-up TDEs, the large majority of events we will detect will be classified as prompt, in which their emission will increase by several orders of magnitude nand then decay over a few weeks to months. However, as highlighted in \citet{2016arXiv161102291A} and references therewithin, it is likely that there is a large population of low luminosity and/or slow rise TDEs that are currently missed or mistaken for other types of events. 

In an attempt to quantify this effect we determined whether each event would have been detectable at various redshifts assuming the flux limit of the \emph{Swift}-BAT, \emph{ROSAT} and \emph{XMM}-slew. For the brightest events (those with a peak X-ray luminosity $\gtrsim 10^{42-43}$ erg s$^{-1}$), \emph{Swift}-BAT would easily detected these sources, while for events with peak luminosities less than this value the \emph{Swift}-BAT would have more difficulty. This is similar for both \emph{ROSAT} and \emph{XMM}-slew. As such, we believe that the brightest sources in our TDE sample are likely to be the most representative type of TDE that emit at these luminosities. However, we acknowledge that our sample unfortunately forms an incomplete picture of the type of events that have peak luminosities $\lesssim 10^{42}$ erg s$^{-1}$. As a consequence the LF that we derive for X-ray luminosity $\gtrsim 10^{42-43}$ erg s$^{-1}$ is more likely to be a more reliable representation of the current soft LF of TDEs than that derived at low luminosities.

Using the full light curves of the X-ray TDE and likely X-ray TDE candidates from \citet{2016arXiv161102291A}, we derive the soft X-ray LF of TDEs in a given luminosity bin, $\Phi(L_{X})$, using the $1/V_{\rm max}$ method developed by \citet{1968ApJ...151..393S}. This method calculates the number density of TDEs in a given luminosity bin, weighted by the maximum volume ($V_{\rm max}$) that each event could have been detected in. In its simplest form, $\Phi(L_{X})$, can be found by summing $1/V_{\rm max}$ for all $N$ TDEs in each luminosity bin. Here, $V_{\rm max}$ was assumed to be our co-moving volume, and was truncated at the maximum volume (redshift) in which we detected an event in our current TDE sample. In Figure \ref{XLF} we have plotted the X-ray LF we derived for all TDEs and for non-jetted TDEs. Overlaid in purple, is the theoretical X-ray luminosity function of TDEs derived by \citet{2006ApJ...652..120M}, while in various shades of colours we have plotted the soft X-ray LF for AGN in different redshift shells as derived by \citet{2005A&A...441..417H}. 

One can see that for luminosities $\lesssim10^{44}$ erg s$^{-1}$, emission from both jetted and non-jetted events contribute non negligibly to the LF of AGN, especially at low redshifts ($z<0.2-\sim0.4$). For higher redshifts, the contribution of X-ray TDEs to the LF of AGN becomes less significant. As theoretically predicted by \citet{2006ApJ...652..120M}, the LF for non-jetted TDEs dramatically falls off around $\sim10^{44}$ erg s$^{-1}$ which corresponds to the Eddington luminosity of a $10^{6}M_{\odot}$ BH. As such, above this luminosity, non-jetted TDEs do not contribute significantly to the X-ray LF of AGN. Due to the super-Eddington nature of jetted TDEs, we find that these events can contribute significantly to the $z\lesssim0.4$ AGN LF, however beyond this redshift, their contribution seems less important. 

The flattening of the TDE LF between $10^{40}-10^{42}$ erg s$^{-1}$ is most likely a consequence of the lack of TDE sources which have a soft X-ray light curve that peaks within this luminosity band (see Figure \ref{lumredvshr}), rather than a real feature. For example, the veiled X-ray TDEs listed in \citet{2016arXiv161102291A} could potentially populate this region of parameter space and naturally explain why we currently see this deficit. As such, we expect that with the detection of more, lower luminosity or weakly emitting X-ray TDEs in the future, the X-ray luminosity function in this region will similarly trace the theoretical expectation of \citet{2006ApJ...652..120M}. Due to the limitations in sensitivity of current X-ray instruments, and the fact that the majority of AGN are found at higher redshifts compared to some TDEs in our sample, current AGN samples are flux limited to a $L_{X}\sim10^{42}$ erg s$^{-1}$. As a consequence, our understanding of the AGN LF below this completeness limit is not well known. Theoretically, the AGN LF should have a similar increasing behaviour as that seen for TDEs, as such we expect that TDEs would continue to contribute significantly at lower luminosities for a range of redshifts.

It is interesting to note just how close the X-ray TDE LF we derive from observations matches the theoretical expectations of \citet{2006ApJ...652..120M}. On the surface this may not look surprising, however \citet{2006ApJ...652..120M} assumed a significantly higher rate of stellar disruptions ($7\times10^{-4}$~yr$^{-1}$ per galaxy) compared to that currently inferred from observational studies in optical/UV and X-ray wavelengths \citep[e.g.,][inferred a rate between $\sim(0.1-1.7)\times10^{-4}$ yr$^{-1}$ per galaxy, however more recently \citet{2017arXiv170300965B} suggest an even higher rate]{2008ApJ...676..944G, 2014ApJ...792...53V, 2016MNRAS.455.2918H}. Our derived TDE LF seems to imply that observationally, the rate of TDEs we are detecting in X-rays is converging towards the theoretically expected rate of tidal disruptions.

Using the BH mass estimates of the X-ray TDE and likely X-ray TDE sample derived by \citet{2016arXiv161102291A}, we can calculate the TDE rate implied by this sample using the relationship between the TDE rate and the mass of the SMBH for both core and cusp galaxies as derived by \citet{2016MNRAS.455..859S} (see Equation 27 of this paper). From this simple exercise, we find that the implied TDE rate is $(0.7-4.7)\times10^{-4}$  yr$^{-1}$ per galaxy, which on the high end of the estimate is similar to the rate used by \citet{2006ApJ...652..120M} than that implied from other observational studies. 

It is possible that the discrepancy we see between the theoretical and observational TDE rates could arise from the fact that many theoretical estimates truncate the BH mass function below $\lesssim10^{6}M_{\odot}$, while the black hole estimates of \citet{2016arXiv161102291A} suggest that a large number of X-ray TDEs have a BH mass between $(10^{5}-10^{6})M_{\odot}$. As TDEs arising from low mass BHs tend to have lower peak luminosities than TDEs from high mass BHs, it is easier for current surveys/observations to detect these events at lower redshift. As such, if the volume in which one detect these events is over-estimated, as would be done if one assumed that TDEs tend to come from larger mass BHs, this would dramatically under-estimate the volumetric rate of TDEs \citep[e.g.,][]{2016MNRAS.461..371K}. Another possibility is that current optical/UV surveys are better suited to detected prompt, rapidly accreting TDEs and are missing a large number due to their viscously delayed nature \citep{2016arXiv161102291A}.

Currently, the observed evolution of the AGN LF with redshift has lead various authors to suggest the low-luminosity end of the AGN LF in different redshift shells is dominated by high-mass BHs with quiescent accretion \citep[e.g.,][]{2005ApJ...630..716H, 2007A&A...474..755B, 2012MNRAS.419.2797F}. However, the significant contribution of TDEs to the LF at these lower luminosities as seen in Figure \ref{XLF} could suggest that TDEs contribute significantly to the growth of BHs with a mass $<10^{7}M_{\odot}$, through close to or super Eddington accretion. Studies suggest that the fraction of galaxies hosting AGN increases with stellar mass \citep{2010ApJ...720..368X}. As such it might be not so surprising that compared to AGN, TDEs could contribute significantly to the growth of BHs with a mass $<10^{7}M_{\odot}$, as AGN activity in these lower mass galaxies is much more repressed compared to that seen in more massive galaxies.

To determine how significantly TDEs contribute to the growth of lower mass BHs, we estimate the mass accretion rate of TDEs using the X-ray LF for non-jetted TDEs seen in Figure \ref{XLF}. Here we assume that 10\% of the mass accreted onto the BH is converted into luminosity. We do not use the TDE LF that includes the jetted events as their beaming fraction is quite uncertain, which can lead to over-estimating the accretion rate in these systems. Multiplying this by the Hubble time, we find that the total mass gained by a BH via the tidal disruptions of stars is significantly less than that expected to be contributed by accretion via an AGN at a comparable redshift. Similar to \citet{2016MNRAS.461..371K}, we find that the total mass gained by a TDE is $\sim10^{5}M_{\odot}$, thus we do not expected TDEs to contribute significantly to the growth of  $10^{6-7}M_{\odot}$ BHs. However, we should note that this might change if one includes the jetted events in this calculation, since they may have a significantly higher mass accretion efficiency compared to that assumed for non-jetted events. This is consistent with both the argument presented by \citet{1982MNRAS.200..115S} that SMBHs grow predominantly via gas accretion and the fact that the AGN LF is well matched by BHs growing via gas accretion \cite[e.g.,][]{2005ApJ...630..716H, 2017MNRAS.tmp..223S}.

However, as we expect the total mass gained by a BH from a TDE to be $\sim10^{5}M_{\odot}$, for BHs with a mass $\lesssim10^{6}M_{\odot}$ it is possible that TDEs could contribute significantly to the growth of these BHs \citep{2009ApJ...697L..77R}. Here one would have to assume that the TDE rate is similar to that expected for $10^{6}M_{\odot}$ BHs. This assumption is not unreasonable as the nuclear cluster immediately surrounding the BH can be replenished with stars over a Hubble time due to the relatively short relaxation timescales of these BHs \citep[e.g.,][]{2005ApJ...621L.101M, 2007ApJ...671...53M}. \citet{2015ApJ...809..166G} showed that half of TDEs caused by a $10^{5-6}M_{\odot}$ BHs will exhibit super-Eddington accretion rates. As these TDEs will be faint (unless they produce a jet) in comparison to their mass accretion rates, these super-Eddington TDEs might contribute significantly to the growth of these lower mass BHs, without contributing to the AGN LF or violating \citet{1982MNRAS.200..115S} argument.

\section{Summary}

In this paper, we have identified some characteristics that help distinguish AGN from TDEs using X-rays. Using the 13 X-ray TDE and likely X-ray TDE candidate classified by \citet{2016arXiv161102291A}, we study the differences in the X-ray emission properties of these events and those of 10 highly variable AGN and AGN detected in deep extragalactic X-ray surveys.

We find that both jetted and non-jetted X-ray TDEs are significantly brighter, with non-jetted (jetted) events much softer (harder) at peak than AGN found at similar redshifts. Even though it was suggested that a non-neglible fraction of AGN detected in extragalactic X-ray surveys may be emission from a TDE, we suggest that luminous TDEs like those in our sample would easily be differentiated from the general AGN population found in these surveys. However, we cannot rule out that low-luminosity TDEs are not currently being misidentified as AGN in these same surveys.

We attempted to differentiate between emission arising from a TDE and that of an AGN by studying how luminous these events become compared to pre-flare constraints, and over short timescales. We find that highly variable AGN are able to produce flare-like emission that exhibits an increase in luminosity of a similar order of magnitude as that seen from X-ray TDEs. However, the emission from a TDE decays significantly more coherently than the emission arising from an AGN.

In addition, we study how the spectral hardness of these events evolve as a function of both time and luminosity. We find that emission from TDEs show little to no variation in their hardness ratio during the decay of a flare, in contrast to the wild variation seen in AGN. We suggest that the lack of variation in HR seen from TDEs could result from inherent difference in the accretion processes (i.e., the stochastic variation in accretion for AGN are much larger than those seen in TDEs) that result in the flare emission from these events, or from the significantly different environments surrounding the central BH. We also find that AGN are significantly more absorbed and produce X-ray emission that is significantly harder than that of TDEs. 

Using our TDE sample, we also derive the soft X-ray LF of TDEs. We find that both jetted and non-jetted events contribute significantly to the LF of AGN at low redshifts ($z\lesssim0.4$). As predicted by \citet{2006ApJ...652..120M}, we find that LF of TDEs dramatically falls off around the Eddington luminosity of a $10^{6}-10^{7} M_{\odot}$ BH. However, above this luminosity jetted TDEs begin to contribute. Interestingly, we find that even though \citet{2006ApJ...652..120M} assumed a much higher TDE rate than currently inferred from observations, our X-ray LF of TDEs matches closely the theoretical LF derived by \citet{2006ApJ...652..120M}. Using the relationship between TDE rate and BH mass derived by \citet{2016MNRAS.455..859S}, and the BH mass estimates of our TDE sample from \citet{2016arXiv161102291A} we estimate a TDE rate of $(0.7-4.7)\times10^{-4}$ yr$^{-1}$ per galaxy, much closer to the theoretically expected rates of TDEs. 

We find that TDEs do not contribute significantly to the growth of $10^{6-7}M_{\odot}$ BHs. However, TDEs arising from super-Eddington accretion, can contribute a non-negligible mass to BHs $<10^{6}M_{\odot}$ without contributing significant X-ray emission to the AGN LF at $z=0$.

\acknowledgements
The authors would like to thank Jamie Law-Smith, Sjoert Van Velzen and the anonymous referee for their helpful and constructive comments which helped improve the quality of the manuscript. E.R.R. is grateful for support from the Packard Foundation, NASA ATP grant NNX14AH37G and NSF grant AST-1615881.

\bibliography{reference}

\end{document}